\documentclass[preprintnumbers,amsmath,amssymb,floatfix,11pt,prd,onecolumn,
superscriptaddress,nofootinbib]{revtex4}
\usepackage{latexsym,float}
\usepackage{epsfig}
\usepackage{epstopdf}
\usepackage{caption}
\usepackage{subfig,color}
\usepackage{amssymb}
\usepackage{amsmath, hyperref}
\begin{document}

\title{\bf Cosmological wormholes in $f(R)$ theories of gravity}

\author{Sebastian Bahamonde}
\email{sebastian.beltran.14@ucl.ac.uk}\affiliation{Department of Mathematics,University College London,
    Gower Street, London, WC1E 6BT, UK}

\author{Mubasher Jamil}
\email{mjamil@sns.nust.edu.pk}\affiliation{Department of Mathematics, School of Natural
    Sciences (SNS), National University of Sciences and Technology
    (NUST), H-12, Islamabad, Pakistan}

\author{Petar Pavlovic}
\email{petar.pavlovic@desy.de}\affiliation{Institut f\"{u}r Theoretische Physik,
    Universit\"{a}t Hamburg, Luruper Chaussee 149, 22761 Hamburg, Germany}

\author{Marko Sossich}
\email{marko.sossich@net.hr}\affiliation{Department of Physics, Faculty of Electrical Engineering and Computing, University of Zagreb,
    Unska 3, HR-10 000 Zagreb, Croatia}

\begin{abstract}
{\centering \bf Abstract:} Motivated by recent proposals of possible wormhole existence in galactic halos, we analyse the cosmological evolution of wormhole solutions
in modified $f(R)$ gravity. We construct a dynamical wormhole that asymptotically approaches FLRW universe, with supporting material going to the perfect isotropic fluid
described by the equation of state for
radiation and matter dominated universe respectively. Our analysis is based on an approximation of a small wormhole - a wormhole that can be treated as matched
with the FLRW metric at some radial coordinate much smaller than the Hubble radius, so that cosmological boundary conditions are satisfied. With a special interest
in viable wormhole solutions, we refer to the results of reconstruction procedure and use $f(R)$ functions which lead to the experimentally confirmed $\Lambda$CDM expansion
history of the universe.
Solutions we find imply no need for exotic matter near the throat of considered wormholes,
while in the limit of $f(R)=R$ this need is always present  during radiation and matter dominated epoch. 
\end{abstract}

 \maketitle

\section{Introduction}
The notion of Lorentzian wormholes (or Morris-Thorne wormholes or simply WH) arose when Morris and Thorne explored the possibility of time travel for humans using
the principles of general relativity (GR) \cite{moris}. Einstein's theory of GR predicts that the structure and geometry of spacetime in the presence of matter is not rigid
but is elastic and deformable. The more compact the object is, the more strong the curvature of space is, which essentially leads to the idea of black holes. However in the
later case, the fabric of spacetime loses its meaning at the curvature singularity. If somehow the formation of singularity is avoided than it would be possible to travel in
and out of the throat, so that there is no restriction to observer's motion on the manifold.
The possibility of such solution to the Einstein field equations was for the first time explored by Flamm \cite{Flamm} recently after the discovery of the GR, but it was later
shown that his solution was unstable. First more detailed analysis of the wormhole solution was done later by Einstein and Rosen \cite{Einstein}.

A typical wormhole is a tube-like structure which is asymptotically flat from both sides. The radius of the wormhole's throat could be a constant or variable depending on its
construction and is termed static or non-static respectively.  GR predicts that to form a WH, an exotic form of matter (violating the energy conditions) must be present near the
throat of the WH. The problem is the dearth of a reasonable source that sustains the wormhole geometry. One possible candidate is the phantom energy (which is a cosmic dynamical
scalar field with negative kinetic energy in its Lagrangian) and is the primed candidate of explaining cosmic accelerated expansion as well \cite{phantom}. Since the existence of
phantom energy is questionable and no other suitable exotic matter candidate is available, an alternative approach is commonly followed: investigation if the modifications of laws
of gravity (i.e. GR), proposed primarily for explanation of accelerated expansion and avoiding singularities, can support the WH geometries. Since the WH is a non-vacuum solution of
Einstein field equations, the presence of some form of energy-matter is necessary to construct a WH. In the framework of modified gravity, the matter content is assumed to satisfy the
energy conditions near the WH's throat, while the higher curvature correction terms in the Lagrangian are required to sustain the WH geometry.

In recent years, theories of modified gravity have got enormous attention to model cosmic accelerated expansion; explaining flat rotation curves of galaxies; wormhole formation and
other esoteric phenomenon near black holes \cite{reviews}. A well-known theory of modified gravity is the $f(R)$ gravity, where $R$ is the Ricci scalar. Simple idea on which this
theory is based is generalizing the integral of action for GR, so that the Ricci curvature scalar is replaced by some arbitrary function, $f(R)$. Field equations obtained in this
fashion have higher degree of complexity, and admit richer set of solutions than the standard GR. Harko and collaborators \cite{harko} constructed solutions of static wormholes
threaded by ordinary matter (satisfying the energy conditions) in $f(R)$ gravity, whereby the curvature/gravitational fluid supports the nonstandard wormhole geometries. Note
that the gravitational fluid cannot be considered as exotic. DeBenedictis and collaborators also obtained new static wormhole solutions in the power-law $R^m$ gravity \cite{fur}.
Rahaman et al \cite{r} obtained new static wormholes in $f(R)$ theory using the non-commutative geometry. It was also demonstrated that viable $f(R)$ theories of gravity, obtained
by demanding the consistency with observational data for the Solar System and cosmological evolution, admit existence of wormhole solutions that do not require exotic matter \cite{viable}. Additionally, in \cite{Myrzakulov:2015kda} it was found a traversable wormhole solution in the framework of mimetic $f(R)$ gravity.
Besides, the static wormhole geometries have also been theorized within curvature-matter coupling theories such as $f(R,T)$, where $T$ is the trace of stress-energy-momentum
tensor \cite{azizi}. Additionally, wormhole geometries have also been explored in generalized teleparallel gravity \cite{tg}, Gauss-Bonnet gravity \cite{gb}
and Lovelock-Brans-Dicke gravity \cite{lbd}. In addition, wormholes minimally coupled to pions supported by a negative cosmological constant were obtained in Ref.~\cite{Ayon-Beato:2015eca}.
Spherically symmetric evolving wormholes were studied in the context of standard GR in \cite{trobo}, and similar analysis was also extended to the case of rotating axially symmetric
wormholes in GR \cite{teo}. It was later shown that the violation of the Null Energy Condition is also necessary in non-symmetric and time-dependent wormholes \cite{Hochberg}.
Some specific solutions for dynamical wormholes were also investigated in $f(R)$ gravity \cite{saeidi}, as well as $f(T)$ theories of gravity \cite{sharif}.

It was recently proposed by Rahaman and collaborators that galactic halo posses some of the characteristics needed to support traversable wormholes \cite{Rahaman}, which was followed
by the discussion on possibility of their detection \cite{kuh}. If these galactic wormholes exist it is natural to assume that they would be produced in the early universe,
and therefore question of their properties and evolution during the expansion of the universe naturally arises. In any case, even if these proposed galactic wormholes do not
exist, investigation of wormholes in the cosmological context is interesting topic in its own right. Therefore, we are curious in the construction of
evolving WH in $f(R)$ gravity, in the framework of standard cosmological assumptions. By construction, the evolving WH metric
is similar to the FLRW metric, hence the WH expands as the Universe expands. The WH spacetime is threaded by anisotropic matter which asymptotically becomes isotropic,
so to match usual description of cosmological ideal fluid. Since the isotropicity about every point in spacetime implies homogeneity (but not vice versa), the matter
distribution is also homogeneous asymptotically. The asymptotic limit may correspond to the cosmological horizon, but it could also be much smaller than the horizon in
the limit of a small wormhole - with geometry approaching Friedmann-Lema\^itre-Robertson-Walker (FLRW) on some small distance from the wormhole throat.

The plan of the paper is as follows: In Sec. II, we give an overview of the $f(R)$ theory and its field equations for the static wormhole geometry. In Sec. III, we introduce the cosmological wormholes in $f(R)$ gravity. In Sec. IV, we numerically solve the wormhole equations for radiation and matter era and then we analyse the energy conditions. Finally we give the conclusion in Sec. V. We work with units $c=1=G$ and use the signature $(-,+,+,+)$.

\section{Static wormholes in $f(R)$ gravity}
The action in $f(R)$ gravity  reads
\begin{equation}
\mathcal{S}=\frac{1}{2\kappa} \int \sqrt{-g}f(R) d^{4}x + \int \mathcal{L}_{m}d^{4}x\,,
\label{akcija}
\end{equation}
where $\kappa=8\pi G$, $\mathcal{L}_{m}$ is the Lagrangian density for the matter and $f(R)$ is a smooth function which depends on the scalar curvature. We can notice that if we set $f(R)=R$ we recover the Einstein-Hilbert action.\\
By varying this action with respect to the metric we find
\begin{equation}
f_{R}(R)R_{\mu \nu} - \frac{1}{2}
g_{\mu \nu}f(R)-(\nabla_{\mu}\nabla_{\nu}-
g_{\mu \nu}\Box)f_{R}(R) = \kappa T_{\mu \nu} \,,
\label{f(R)eq}
\end{equation}
where $f_{R}(R)=df(R)/dR$ and $T_{\mu \nu}$ is the energy-momentum tensor of the matter.\\
The Morris-Thorne metric which can describe the space-time of a static wormhole is given by \cite{moris}
\begin{equation}
ds^{2}=-e^{2\Phi(r)} dt^{2}+\frac{1}{1-\frac{b(r)}{r}}dr^{2}+r^{2}(d\theta^{2}+\sin^{2}\theta d\varphi^{2}).
\label{Morristhorne}
\end{equation}
where $\Phi(r)$ is the redshift function and $b(r)$ is a shape function, which are both functions of radial coordinate, $r$. In the wormhole geometry,
radial coordinate needs to non-monotonically decrease from infinity to a minimal value $r_{0}$ at the throat, where $b(r_{0})=r_{0}$, and then increase to infinity.
Although metric is singular at $r=r_{0}$, proper distance as an invariant quantity must be well behaved, and therefore the following integral must be real and
regular outside the throat,
\begin{equation}
 l(r)=\pm\int^{r}_{r_{0}}\frac{dr}{\sqrt{1-\frac{b(r)}{r}}}\,,
\end{equation}
and from this one obtains the condition
\begin{equation}
 1-\frac{b(r)}{r}\geq0 \,.
\end{equation}
 Far from the throat, space must be asymptotically flat which implies the condition $b(r)/r\rightarrow0 $ as $ l\rightarrow\pm\infty $.
By definition, throat represent the minimum radius in this wormhole geometry and this  leads to the flaring-out condition \cite{moris},
\begin{align}
\frac{b(r)-b'(r)r}{b(r)^{2}}>0\,.
\end{align}
Hereafter, prime denotes derivative with respect to the argument of a function, so that $b'(r)=db/dr$. It is also assumed that wormhole has no horizons, so that it can be traversable and
for this reason $\Phi(r)$ must be finite everywhere.
As already mentioned in the introduction, static wormholes in Einstein's General relativity require exotic fluid, i.e. fluid
which is violating standard condition on stress-energy tensor, Weak Energy Condition (WEC). This condition is given by $T_{\mu\nu}k^{\mu}k^{\nu}\geq0$ for any spacelike
vector $k^{\mu}$ \cite{haw}. Apart from construction of cosmological wormholes in $f(R)$  gravity, we will also be interested in
whether this wormholes, that could be produced in the early universe and be determined by its dynamics, could satisfy WEC or not. Moreover, we will also consider the limit $f(R)=R$ as a comparison.

\section{$f(R)$ Cosmological Wormholes}
 Static wormhole geometry (\ref{Morristhorne}) can easily be generalised to the evolving case,
\begin{equation}
ds^{2}=-e^{2\Phi(t,r)} dt^{2}+a^2(t)\Big[\frac{1}{1-\frac{b(r)}{r}}dr^{2}+r^{2}(d\theta^{2}+\sin^{2}\theta d\varphi^{2})\Big]\,,
\label{metrika}
\end{equation}
where $a(t)$ is the scale factor which controls the dynamic of the wormhole. Since parameters $b(r)$ and $\Phi(r,t)$
determine the geometrical properties of the wormhole,
we will naturally demand that they obey the same conditions as in the static case with $b\rightarrow a(t) b(r)$ and $r\rightarrow  a(t)r$. We can notice that if we
set $a(t)\rightarrow 1$ and $\Phi(r,t)=\Phi(r)$, we recover
the static Morris-Thorne metric (\ref{Morristhorne}). In addition, for an observer far away from the throat
of the wormhole we demand that $\Phi(t,r)\rightarrow 0$ and $b(r)\rightarrow 0$ fast enough, so that
for the large enough distances the metric converts to the FLRW one. In other words, this metrics allows
wormhole solutions to be approximately embedded in FLRW universes.
This geometry is clearly not asymptotically-flat, and therefore we here use somehow extended definition
of a wormhole which does not imply asymptotic flatness. In any case, this
implies only generalization of the wormhole's asymptotic behavior, and makes no changes in the throat
properties, which is the most interesting and fundamental attribute of a
wormhole. Related question of formal definitions regarding dynamical wormholes, including wormholes which
are asymptotically FLRW, was previously discussed in \cite{trobo,hideki}.
We can choose a radius $r_{c}\ll 1/H$ for all times (where $H=\dot{a}(t)/a(t)$ is the cosmological Hubble parameter)
such that if we take some $r>r_{c}$, the redshift and shape
function can be neglected and we can approximately treat the wormhole as confined within the
region $r<r_{c}$ with no influence on the global geometry of FLRW space-time. Therefore,
we can demand that the dynamical properties of this wormhole,  apart from the evolution of the red-shift function,
are determined only by the
expansion of the universe, so that the scale factor $a(t)$ is equal in both regions. We can
call this approximate geometry - where $r_{c}$ can be effectively treated as a wormhole's
asymptotic infinity - a small cosmological wormhole.
Objects like this could be created on the microscopical scales in the early
universe and  then subsequently enlarged during expansion of the universe. For
instance, if such wormholes were created at or before the electroweak scale their
physical size $a(t)\,r_c $ should be much less than a centimetre, a typical scale of the Hubble radius at
that time. By the end of matter dominated era such wormholes would then be characterized 
by the typical size of astrophysical objects.

Thus, on the cosmological scales,
the energy-momentum tensor is represented with an isotropic ideal fluid with equation of state
\begin{align}
p_{c}(t)&=w \rho_{c}(t)\,,
\end{align}
where $w$ is the state parameter and $\rho_{c}(t)$ and $p_{c}(t)$ are the energy density and pressure of the cosmological fluid respectively. Here, index $c$ denotes
that we are dealing with functions on the cosmological scales of FLRW metric to distinguish them from the components of the anisotropic energy momentum tensor defined
in the region $r<r_{c}$, which we will denote with the subscript ``$wh$''. \\
If we use the conservation law for the energy-momentum tensor for the cosmological fluid, $\nabla_{\mu}T^{\mu\nu}_{(\rm c)}=0$, we find that the energy density of this fluid is given by
\begin{align}
\rho_{c}(t)&=\rho_{0}a(t)^{-3(1+w)}\,,
\end{align}
where $\rho_{0}$ is the current energy density. It is important to mention that in both regions $r>r_{c}$ and $r<r_{c}$, the form of $f(R)$ should naturally be the identical.\\
The $f(R)$ field equations (\ref{f(R)eq}) for the cosmological flat FLRW spacetime ($r>r_{c}$ region) yield
\begin{align}
H(t)^2&=\frac{1}{3f_{R}(R)}\Big(\frac{1}{2}\Big[R_{c}f_{R}(R)-f(R)\Big]-3H(t)\dot{R_{c}}f_{RR}+\rho_{c}(t)\Big)\,,
\label{mod.friedman}
\end{align}
where $R_{c}=6(\dot{H}+2H^2)$ is the scalar curvature of the flat FLRW spacetime, $f_{RR}=d^{2}f(R)/dR^2$ and dots represent differentiation with respect to the cosmic time.

In the region $r<r_{c}$, matter supporting the wormhole will be an anisotropic fluid such as the energy-momentum tensor is
given by $T^{\mu}_{\nu}=\textrm{diag}(-\rho_{wh}(r,t),p_{r_{wh}}(r,t),p_{l_{wh}}(r,t),p_{l_{wh}}(r,t))$, where $\rho_{wh}(r,t)$, $p_{r_{wh}}(r,t)$ and $p_{l_{wh}}(r,t)$
are the energy density, radial pressure and lateral pressure respectively. It follows that for an anisotropic fluid, WEC
is given by the conditions
\begin{align}
\rho_{wh}(r,t) &\geq 0\,,\\
 \rho _{wh}(r,t) + p_{r_{wh}}(r,t) &\geq 0\,,\\
\rho_{wh}(r,t) + p_{l_{wh}}(r,t) &\geq 0\,.
 \end{align}
The $f(R)$ field equations (\ref{f(R)eq}) in this region ($r<r_{c}$) for the non-static metric (\ref{metrika}) are given by
\begin{eqnarray}
-\rho &=&-\frac{1}{2}f+a^{-2}\Big(1-\frac{b(r)}{r}\Big)R'^2f_{RRR}+\displaystyle \frac{f_{R}}{2r^2a^2}\Big[\left(r \left(b'(r)-4\right)+3 b(r)\right) \Phi'-2 r (r-b(r)) \Phi'^2\nonumber\\
&&+2 r (b(r)-r) \Phi''\Big]+\displaystyle\frac{f_{RR}}{2r^2a^2}\Big[2 r (r-b(r)) R''-\left(r \left(b'(r)-4\right)+3 b(r)\right) R'-6 r^2 a \dot{a}\dot{R}e^{-2 \Phi}\Big]\nonumber\\
&&+\frac{3 f_{R} e^{-2 \Phi}}{a} \left(\ddot{a}-\dot{a} \dot{\Phi}\right)\,,\label{fieldeq1a}\\
p_{r}&=&-\frac{1}{2}f+\frac{b(r)f_{R}}{2r^3a^2}\Big[2 r^2 \Phi'^2+2 r^2 \Phi''-r \Phi'-2\Big]+\frac{e^{-2 \Phi}f_{R}}{2r^2a^2}\Big[e^{2 \Phi} \Big(b'(r) \Big(r \Phi'+2\Big)\nonumber\\
&&-2 r^2 \Big(\Phi'^2+\Phi''\Big)\Big)+4 r^2 \dot{a}^2\Big]+\frac{f_{R} e^{-2 \Phi}}{a} \left(\ddot{a}-\dot{a} \dot{\Phi}\right)-e^{-2 \Phi }f_{RRR} \dot{R}^2\nonumber\\
&&+\frac{f_{RR}R'}{a^2}\left(1-\frac{b(r)}{r}\right)\left(\Phi'+\frac{2}{r}\right)+\frac{e^{-2\Phi}f_{RR}}{a}\Big[a \left(\dot{R}\dot{\Phi}-\ddot{R}\right)-2 \dot{a} \dot{R}\Big]\,,\label{fieldeq1b}\\
p_{l}&=&-\frac{1}{2}f+\frac{f_{R}}{2 r^3 a^2}\Big[b(r)\left(2 r \Phi'+1\right)+\left(b'(r)-2 r \Phi'\right)r \Big]+\frac{e^{-2\Phi}f_{R}}{2a^2}\Big[4  \dot{a}^2+2 a \left(\ddot{a}-\dot{a} \dot{\Phi}\right)\Big]\nonumber\\
&&+f_{RRR}\Big[\frac{R'^2}{a^2}\Big(1-\frac{b(r)}{r}\Big)-\dot{R}^2 e^{-2 \Phi}\Big]+\frac{f_{RR}}{2r^2a^2}\Big[R' \left(r \left(2 r \Phi'-b'(r)+2\right)-b(r) \left(2 r \Phi'+1\right)\right)\nonumber\\
&&+2 r (r-b(r)) R''\Big]-\frac{f_{RR}}{r^3a^2}\Big[4 r^3 a \dot{a} \dot{R}-2 r^3 a^2 \dot{R}\dot{\Phi}+2 r^3 a^2\ddot{R}\Big]\,,\label{fieldeq1c}\\
2\Big(\frac{\dot{a} }{a}\Big)f_{R}\Phi'&=&f_{RRR}\dot{R}R'-f_{RR}\Big[\frac{\dot{a}R'}{a}+\dot{R} \Phi'-\dot{R'}\Big]\,,
    \label{fieldeq1d}
\end{eqnarray}
where $f_{RRR}=d^{3}f(R)/dR^{3}$, and $R$ is the Ricci scalar for the evolving wormhole geometry (\ref{metrika}). We will
also consider the following boundary conditions (at the boundary $r=r_{c}$)
\begin{align}
\rho_{wh}(r_{c},t)&=\rho_{c}(t)\,,\\
p_{r_{wh}}(r_{c},t)&=p_{c}(t)\,,\\
p_{l_{wh}}(r_{c},t)&=p_{c}(t)\,.
\end{align}
These conditions imply that the anisotropic fluid supporting the wormhole goes continuously to the isotropic fluid of the Universe. Moreover, they also tell us that the density-pressure
relationship goes to the cosmological equation of state at $r=r_{c}$,
\begin{align}
 p_{r_{wh}}(r_{c},t)=p_{l_{wh}}(r_{c},t)=w \rho_{wh}(r_{c},t)\,.
 \end{align}
There are four field equations (\ref{fieldeq1a})-(\ref{fieldeq1d}) containing seven unknowns
functions, therefore we need to close the system by choosing some ansatz. Hereafter, we
will consider that $\Phi(r,t)=h(r)\phi(t)$ where $h(r)$ will be some chosen function which decreases fast to
negligible values for  $r\geq r_{c}$. It is also necessary to prescribe the form of the modified gravity, $f(R)$, which
from (\ref{mod.friedman}) determines $a(t)$ and therefore the expansion of the small cosmological wormhole. With special interest in viable wormhole
solutions we will consider choices for $f(R)$ which lead to experimentally confirmed $\Lambda$CDM expansion history of the Universe. Thus, this choice of $f(R)$ will
lead to known forms of $a(t)=a_0(t/t_0)^n $ in radiation ($n=1/2$)
and matter dominated epoch ($n=2/3$). Following the results of this reconstruction procedure we take \cite{reconstruction,Nojiri:2006gh,Nojiri:2009kx,Nojiri:2006be,Dunsby:2010wg}
\begin{equation}
f(R)_{\rm rad}= R - 3 D (\kappa \rho_{m}^{0})^{4/3} R^{-1/3},\label{fRad}
\end{equation}
during radiation dominated phase, where $\rho_{m}^{0}$ is the current energy density of matter and $D$ is an arbitrary constant.
For the matter dominated phase we will consider \cite{reconstruction,Nojiri:2006gh,Nojiri:2009kx,Nojiri:2006be,Dunsby:2010wg}
\begin{equation}
f(R)_{\rm mat}=R -\zeta \Big(\frac{R}{\Lambda}\Big)^{1-p}\,,\label{fmatter}
\end{equation}
where we have defined the constant $\zeta$ as
\begin{equation}
\zeta  = \frac{240 D\, \Gamma(\frac{\sqrt{73}}{2})}{\Gamma (\frac{7+ \sqrt{73}}{4}) \Gamma (\frac{13+ \sqrt{73}}{4})}
\kappa^{2} \rho_{r}^{0} \frac{(R_{0} - 4\Lambda)}{\Lambda^{p-1}} (\frac{\rho_{m}^{0}}{\rho_{r}^{0}})^{3(p-1)}\,,
\end{equation}
where $\rho_{r}^{0}$ is the current energy density of radiation and $p=(5+\sqrt{73})/12\approx 1.13$.
It can be shown that $f(R)_{\rm rad}$ and $f(R)_{\rm mat}$ are the special limits of one more general $f(R)$ form in terms of hypergeometric functions, capable of
describing both radiation and matter dominated era \cite{reconstruction}. 

\section{NUMERICAL SOLUTIONS}

\subsection{Radiation dominated era }
In this section, we will consider the evolution of cosmological wormhole during the radiation dominated era, where cosmological perfect
fluid is described by the EOS parameter $w=1/3$, leading to the expansion of the scale factor as $a(t)=a_{0}(t/t_{0})^{n}$. We will consider a small wormhole for
which $r_{c}=10 \times r_{0}$ with $r_{0}<<1/H$ for all times. We choose the shape function to be
$b(r)=r_{0}^{m+1}/r^{m}$, and take $h(r)= \varphi_{0} e^{(-r/r_{0}+1)^{n}}$ where $m$ and $n$ are integers. Also, we take $D=(1/3)(\kappa \rho^{0}_{m})^{-4/3}$ and parameters
for the energy densities today are set according to the standard cosmological values \cite{planck}. Our results will be presented until
the end of radiation dominated epoch, $t/t_{0}=10^{-6}$, where $t_{0}$ is the time passed from the Big Bang until now.

We first consider the non-diagonal modified Einstein's equation (13), which is a partial differential equation with respect to $\Phi(r,t)$ (note that Ricci curvature
is itself dependent on $\Phi(r,t)$, so this equation contains derivatives with respect to both the radial and time coordinate, as well as the mixed terms). In order to solve this
equation we will solve it at the throat, $r=r_{0}$, and assume that the same time dependence will remain valid for all other positions. This is consistent with the
approximation that $\Phi(r,t)$ goes fast to zero for $r>r_{0}$, so that only important region is that in the vicinity of the throat. After $\phi(t)$ has been numerically obtained in this
manner 
, it can be used to solve the remaining modified Einstein's equation, which represent only algebraic equations for the
components of energy-momentum tensor of anisotropic fluid. In Figs.~\ref{Fig2}, \ref{Fig3} and \ref{Fig4}  we show the obtained
 time evolution and the space dependence of $\rho_{wh}(r,t)$, and check for the possible WEC violation related to both pressures:
$\rho_{wh}(r,t) + p_{r_{wh}}(r,t)$ and $\rho_{wh}(r,t) + p_{l_{w}h}(r,t)$. We also show $w_{r_{wh}}=p_{r_{wh}} /\rho_{wh} $ (see Fig.~\ref{Fig7}) as well as the difference
between radial and lateral pressure, $p_{r_{wh}} - p_{l_{wh}}$, as
a natural measure of anisotropicity of the fluid (see Fig.~\ref{Fig6}). One can clearly see that the matter supporting the wormhole goes to the ideal isotropic cosmological
fluid with $w_{r_{wh}}=w_{l_{wh}} \approx 1/3$ when
approaching $r_{c}$. Moreover, WEC is always satisfied, so there is no need for introducing any form of exotic matter to
support the wormhole. The exoticity parameter $\xi$ describes the physical nature of matter near and far away from the wormhole's throat. If $\xi>0,(\xi<0)$ than
the matter surrounding the wormhole is exotic (non-exotic). From
Fig.~\ref{Fig7b}, it is obvious that $\xi<0$ for all values of $r$, thus
matter (in this case, radiation) remains non-exotic and the wormhole
is stable even with radiation and do not require exotic matter.

In the region  $r>r_{c}$ flat FLRW metric is by construction replacing the dynamical wormhole metric to which it is connected, leading to the known
solutions that we do not plot here. 
In the shown plots we choose $t/t_{0}=10^{-9}$ as a suitable origin, but
 with the presented conclusions remaining valid for the earlier times after the Big Bang.
The last issue that then remains to be discussed is the exact numerical establishment of the boundary condition $\rho_{wh}(r_{c},t)=\rho_{c}(t)$. This can be achieved by
choosing the appropriate values
for the free parameters $m$,$n$, $\phi_{0}$ and $ \dot{\phi}(t_{0})$. Since this adjusting of the free parameters is not very enlightening, for the sake of
the simplicity we choose to demonstrate this in a different manner. We absorb the dependence on the free parameters in a constant term, $C$, which is
multiplying our solution for energy density, $\rho_{wh}(r_{c},t)$, obtained with arbitrary set of parameters. It can be seen in Fig.~\ref{Fig8} that with choosing the different
constant parameters one obtains the family of curves with the same evolution as the cosmological density, $\rho_{c}(t)$. For a suitable value of $C$ curves $\rho_{wh}(r_{c},t)$ and
$\rho_{c}(t)=\rho_{0}a^{-4}$ exactly coincide. \\
\begin{figure}[H]
	\captionsetup{justification=raggedright
	}
    \centering
    \includegraphics[width=0.46\textwidth]{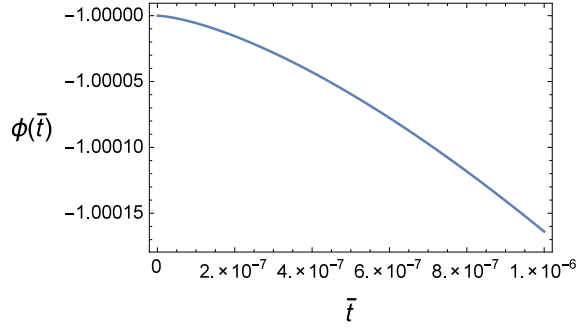}
    \caption{\scriptsize{Time-dependent part of the redshift function $\phi(t)$ as a function of the dimensionless time $\bar{t}=t/t_{0}$ for a radiation dominated era with a
    shape function $b(r)=r_{0}^{m+1}/r^{m}$ and $h(r)=\varphi_{0}e^{(-r/r_{0}+1)^n}$. This pictures represents a slice $r=r_{0}$ with
     $m=2$, $n=1/2$, $\phi(\bar{t}=10^{-9})=-1$ , $\dot{\phi}(\bar{t}=10^{-9})=-10$ and $\varphi_{0}=1$}}\label{Fig1}
\end{figure}

\begin{figure}[H]
		\captionsetup{justification=raggedright}
    \subfloat[Energy density]{\includegraphics[width=0.5\textwidth]{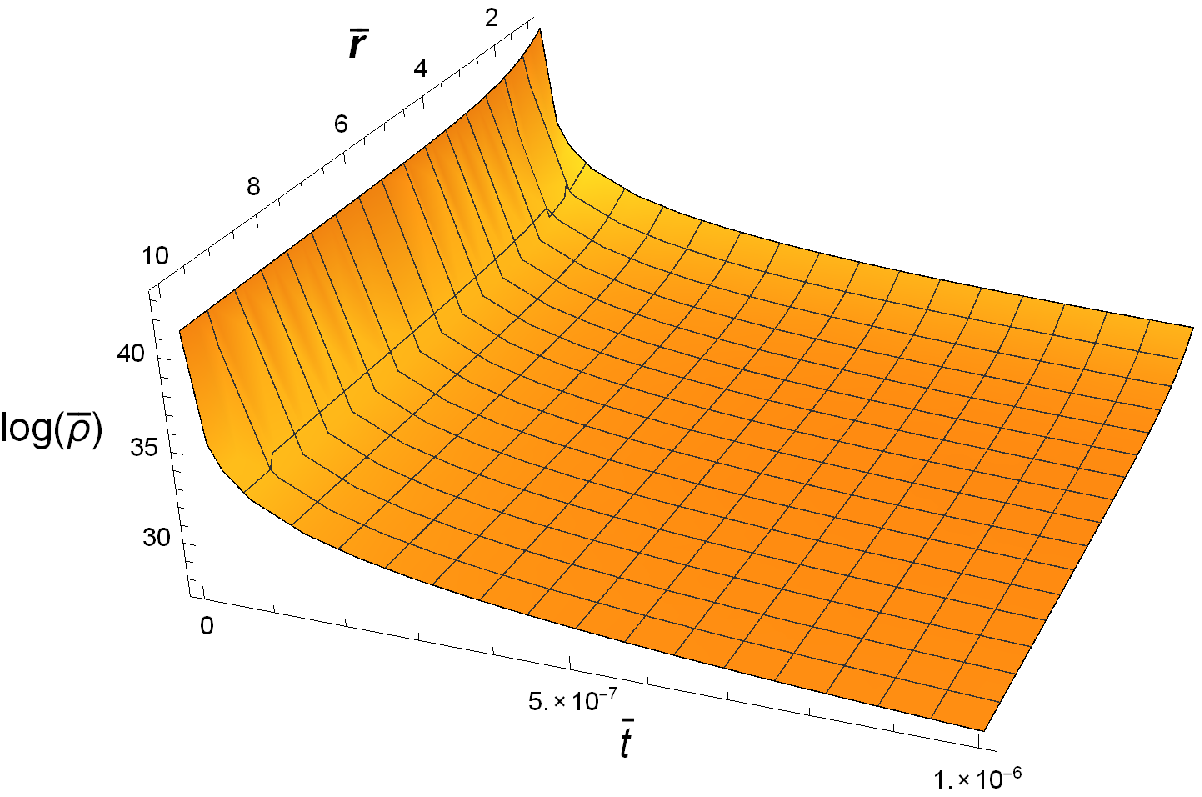}\label{Fig2}}
     \hfill
    \subfloat[Pressure difference]{\includegraphics[width=0.5\textwidth]{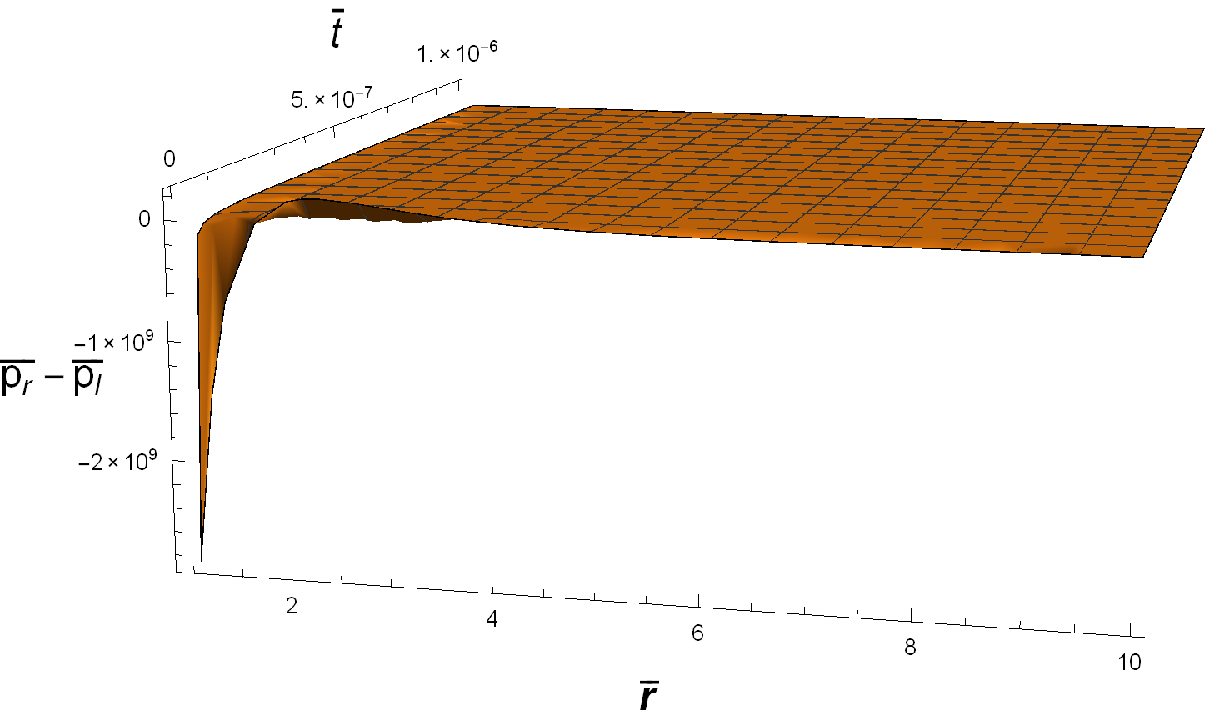}\label{Fig6}}
    \hfill
    \subfloat[WEC-1]{\includegraphics[width=0.5\textwidth]{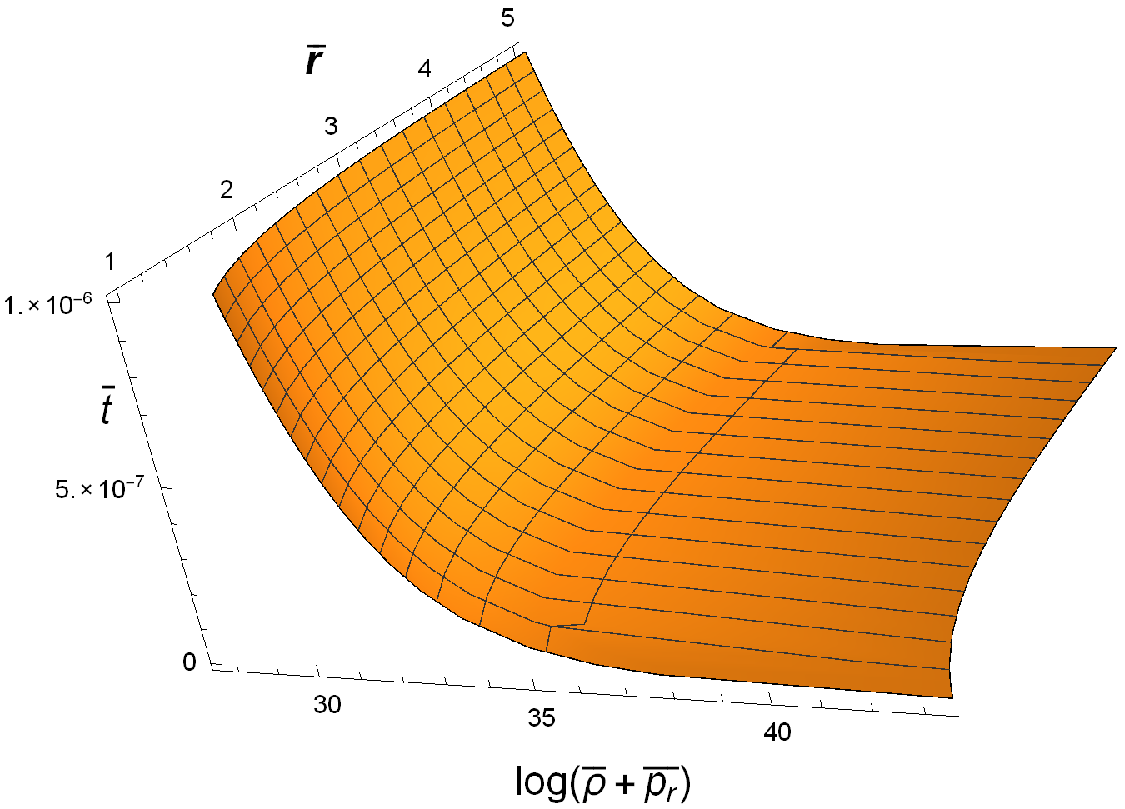}\label{Fig3}}
       \subfloat[WEC-2]{\includegraphics[width=0.5\textwidth]{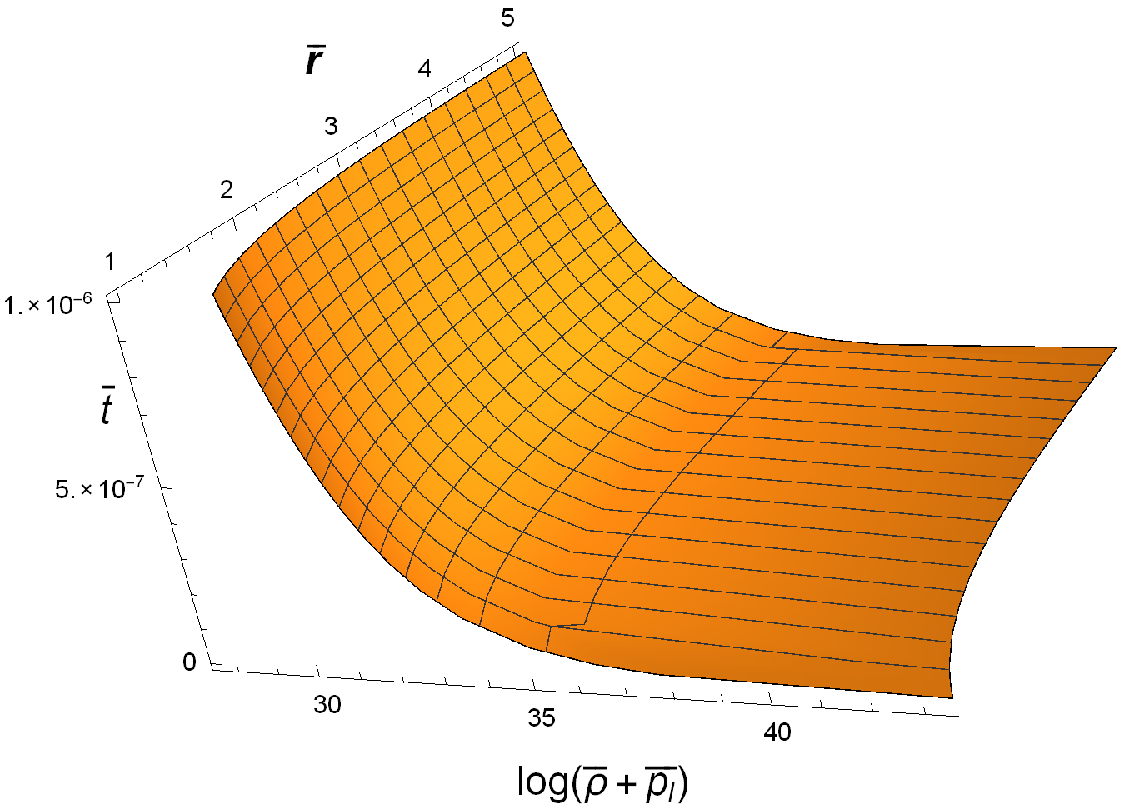}\label{Fig4}}
    \caption{\scriptsize{Figures show logarithmic of the dimensionless energy density $\bar{\rho}=\rho/\rho_{c}$,
    difference in dimensionless pressures $\bar{p}_{l}(t,r)-\bar{p}_{r}(t,r)$, logarithmic
    WEC-1 ($\bar{p}_{r}+\bar{\rho}$) and logarithmic WEC-2 ($\bar{p}_{l}+\bar{\rho}$) as a
    function of the dimensionless time $\bar{t}=t/t_{0}$ and the dimensionless radius $\bar{r}=r/r_{0}$
    respectively. The parameters used were $m=2$, $n=1/2$, $\rho^{0}_{m}=0.27$, 
         $\varphi_{0}=1$ and $t/t_{0}=10^{-9}$ as the origin with a critical
         density $\rho_{c}=3/(\kappa t_{0}^2)$, a shape function $b(r)=r_{0}^{m+1}/r^{m}$, a
         redshift function given by $\Phi(t,r)=\varphi_{0}e^{(-r/r_{0}+1)^n}\phi(t)$ (with $\phi(t)$
         being displayed in Fig. (\ref{Fig1}) and the function $f(R)$ given by (\ref{fRad})).}}
\end{figure}

\begin{figure}[H]
		\captionsetup{justification=raggedright
		}
    \centering
    \subfloat[WH state parameter]{\includegraphics[width=0.45\textwidth]{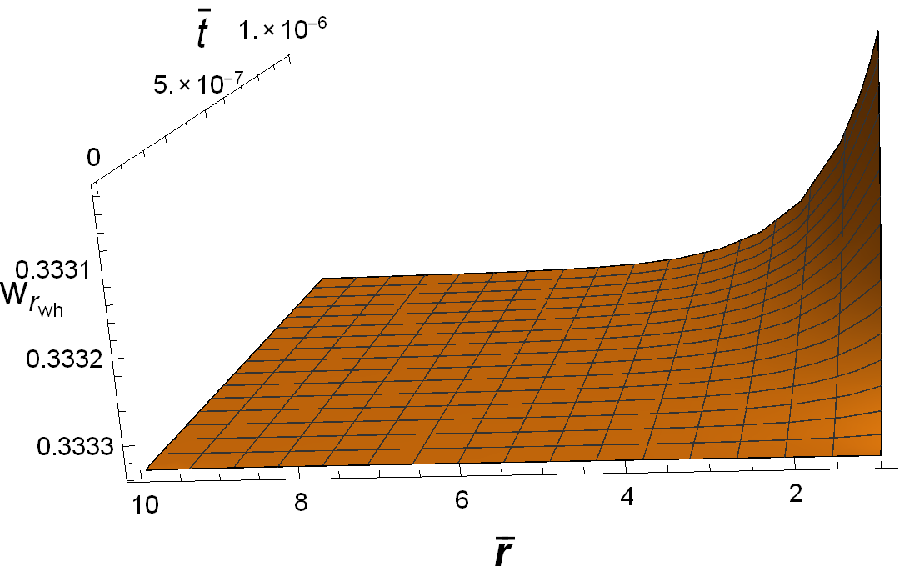}\label{Fig7}}
    \hfill
    \subfloat[Exoticity]{\includegraphics[width=0.45\textwidth]{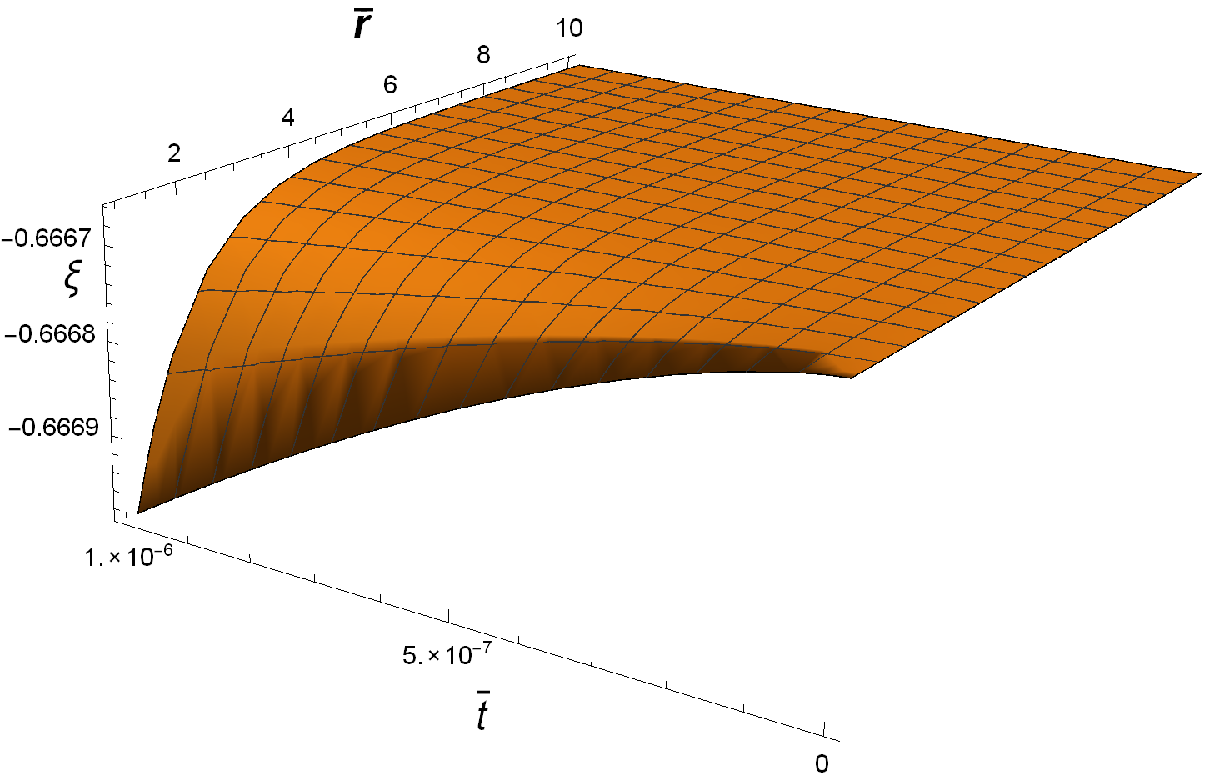}\label{Fig7b}}

   \caption{ \scriptsize{Figures show the wormhole state parameter $w_{r_{wh}}=p_{r_{wh}}/\rho_{r_{wh}}$ and the
        exoticity parameter $\xi=(p-\rho)/|\rho|$ for radiation dominated era as a function of the
        dimensionless time $\bar{t}=t/t_{0}$ and the dimensionless radius $\bar{r}=r/r_{0}$ .  The
        parameters used were $m=2$, $n=1/2$,
        $\varphi_{0}=1$, $\rho^{0}_{m}=0.27$ and  $t/t_{0}=10^{-9}$ as
        the origin with a critical density $\rho_{c}=3/(\kappa t_{0}^2)$, a shape
        function $b(r)=r_{0}^{m+1}/r^{m}$, a redshift function given
        by $\Phi(t,r)=\varphi_{0}e^{(-r/r_{0}+1)^n}\phi(t)$ (with $\phi(t)$ being
        displayed in Fig. (\ref{Fig1}) and the function $f(R)$ given by (\ref{fRad}))}}
    
\end{figure}

\begin{figure}[H]
    \centering  \includegraphics[width=0.75\textwidth]{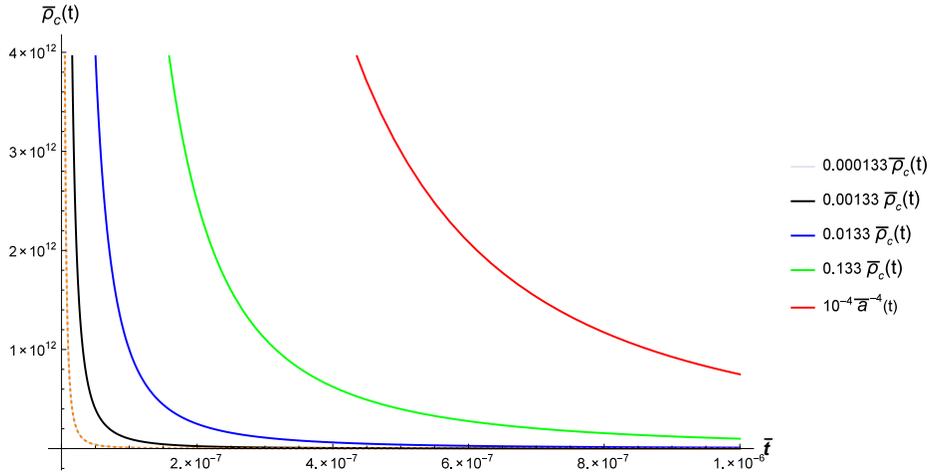}
    \caption{\scriptsize{Figure shows a family of curves with the same evolution as the cosmological dimensionless density  $\bar{\rho}_{c}(t)$ evaluated at $r_{c}=10$ versus dimensionless
    time $\bar{t}=t/t_{0}$. The curve $10^{-4}\bar{a}(t)^{-4}=10^{-4}\bar{t}^{-2}$  is overlaped with the curve $0.000133\,\bar{\rho}_{c}(t)$,
    so that  $C=0.000133$ is a suitable value.
    The parameters used were $m=2$, $n=1/2$,
          $\varphi_{0}=1$, $\rho^{0}_{m}=0.27$ and  $t/t_{0}=10^{-9}$ as
          the origin with a critical density $\rho_{c}=3/(\kappa t_{0}^2)$, a shape
          function $b(r)=r_{0}^{m+1}/r^{m}$, a redshift function
          given by $\Phi(t,r)=\varphi_{0}e^{(-r/r_{0}+1)^n}\phi(t)$ (with $\phi(t)$
          being displayed in Fig. (\ref{Fig1}) and the function $f(R)$ given by (\ref{fRad}))}}\label{Fig8}
\end{figure}

\subsection{Matter dominated era}
\begin{figure}[H]
\centering  \includegraphics[width=0.5\textwidth]{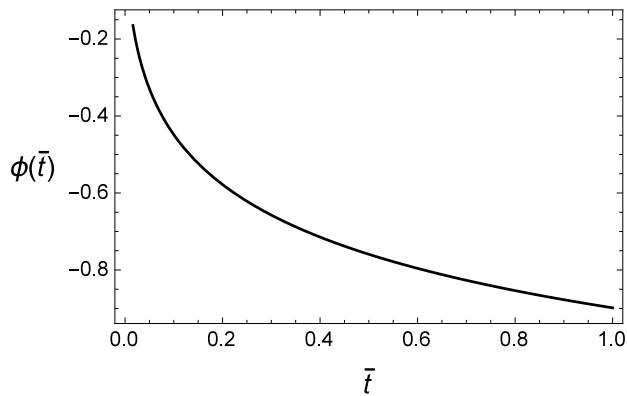}
    \caption{\scriptsize{Time-dependant part of the redshift function $\phi(t)$ versus
    the dimensionless time $\bar{t}=t/t_{0}$ for a matter dominated era with a
    shape function $b(r)=r_{0}^{m+1}/r^{m}$ and $h(r)=\varphi_{0}e^{(-r/r_{0}+1)^n}$.
    This pictures represents a slice $r=r_{0}$ and the parameters
    used was $m=2$, $n=2/3$, $\phi(\bar{t}=10^{-6})=1$ , $\dot{\phi}(\bar{t}=10^{-6})=-1$ and $\varphi_{0}=1$. }}\label{Fig9}
\end{figure}

\begin{figure}[H]
    \centering
    \subfloat[Energy density]{\includegraphics[width=0.4\textwidth]{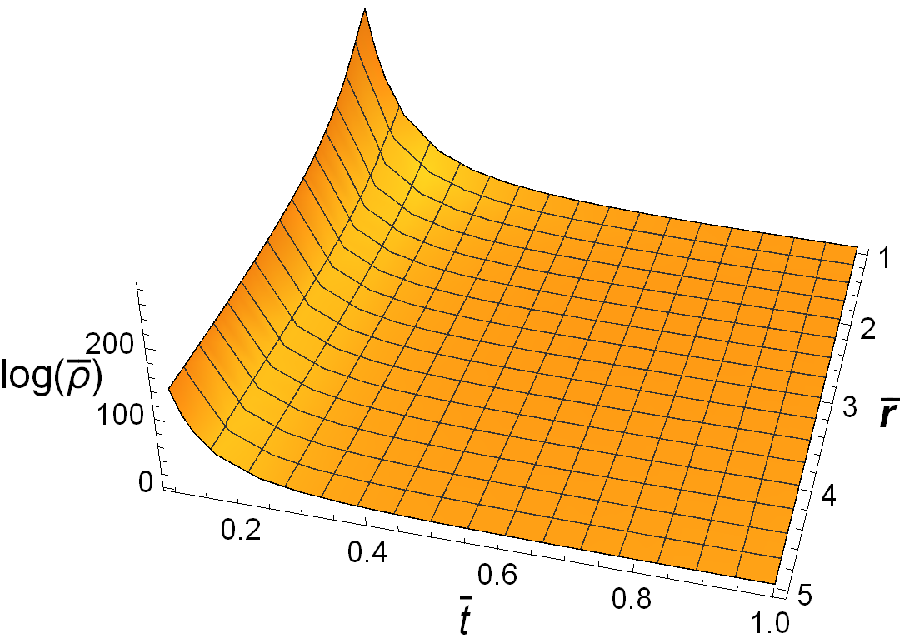}\label{Fig10}}
    \hfill
    \hspace{2mm}
        \subfloat[Pressure difference]{\includegraphics[width=0.5\textwidth]{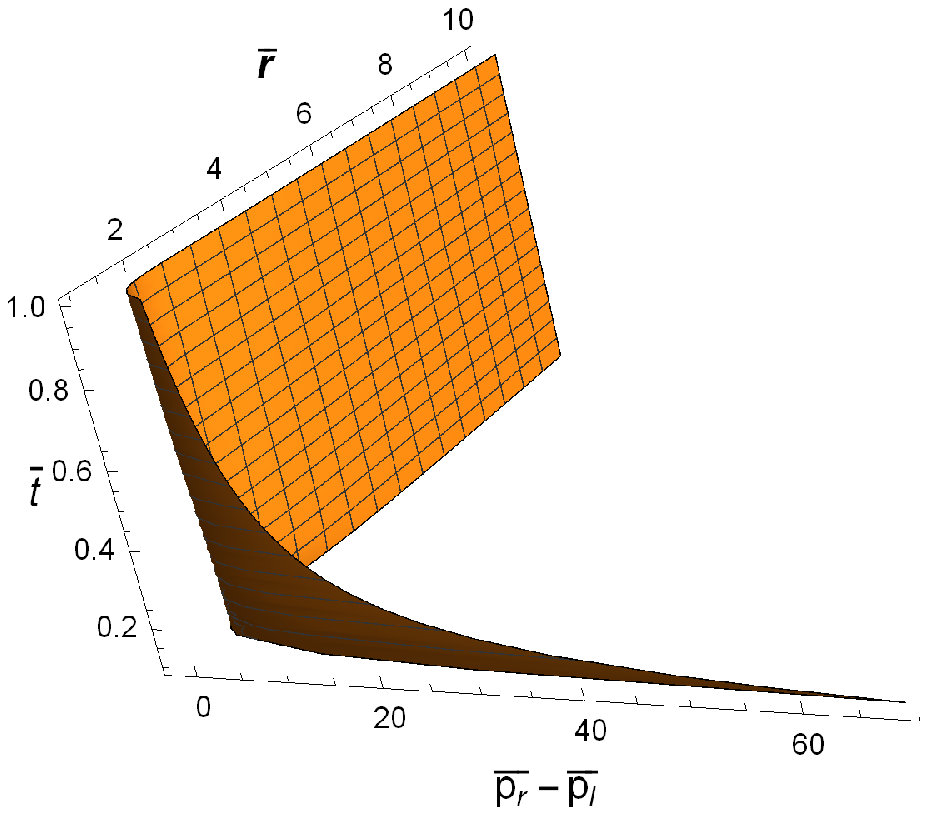}\label{Fig15}}
        \hfill
    \subfloat[WEC-1]{\includegraphics[width=0.5\textwidth]{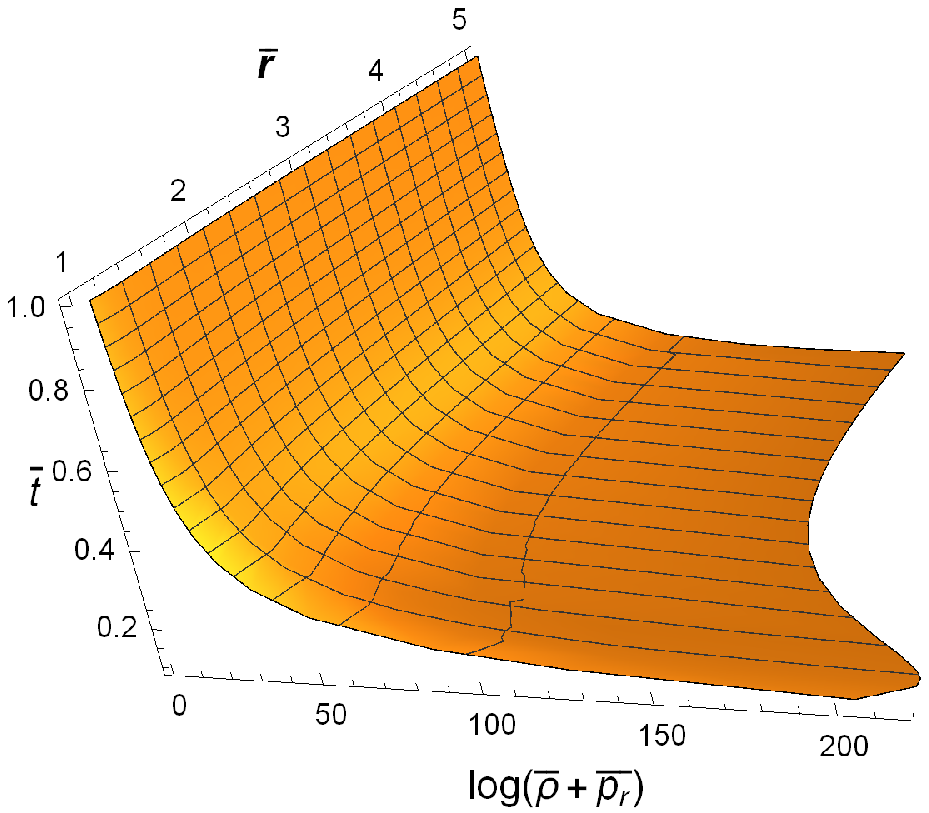}\label{Fig11}}
    \hfill
    \subfloat[WEC-2]{\includegraphics[width=0.5\textwidth]{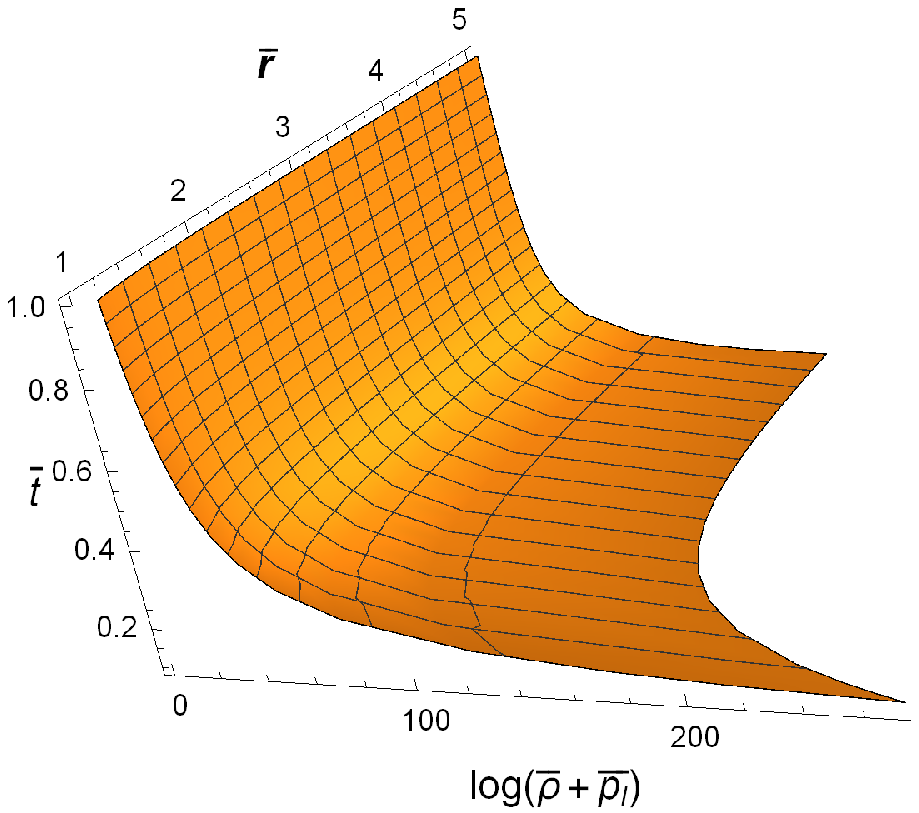}\label{Fig12}}
    \caption{\scriptsize{Figures show the dimensionless energy density $\bar{\rho}=\rho/\rho_{c}$, difference in dimensionless pressures $\bar{p}_{l}-\bar{p}_{r}$, WEC-1 ($\bar{p}_{r}+\bar{\rho}$) and WEC-2 ($\bar{p}_{l}+\bar{\rho}$) as a function of the dimensionless time $\bar{t}=t/t_{0}$ and dimensionless radius $\bar{r}=r/r_{0}$ respectively. In this case, we have used the parameters $m=2$, $n=2/3$, $\Lambda=0.001$, $r_{0}=\varphi_{0}=\zeta=1$, $p=(5+\sqrt{73})/12$,
         and  $t/t_{0}=0.1$ as the origin with a critical density $\rho_{c}=3/(\kappa t_{0}^2)$, a shape function $b(r)=r_{0}^{m+1}/r^{m}$, a redshift function given by $\Phi(t,r)=\varphi_{0}e^{(-r/r_{0}+1)^n}\phi(t)$ (with $\phi(t)$ being displayed in Fig. (\ref{Fig9}) and the function $f(R)$ given by (\ref{fmatter}))} }
\end{figure}

\begin{figure}[H]
    \centering
    \subfloat[WH state parameter]{\includegraphics[width=0.45\textwidth]{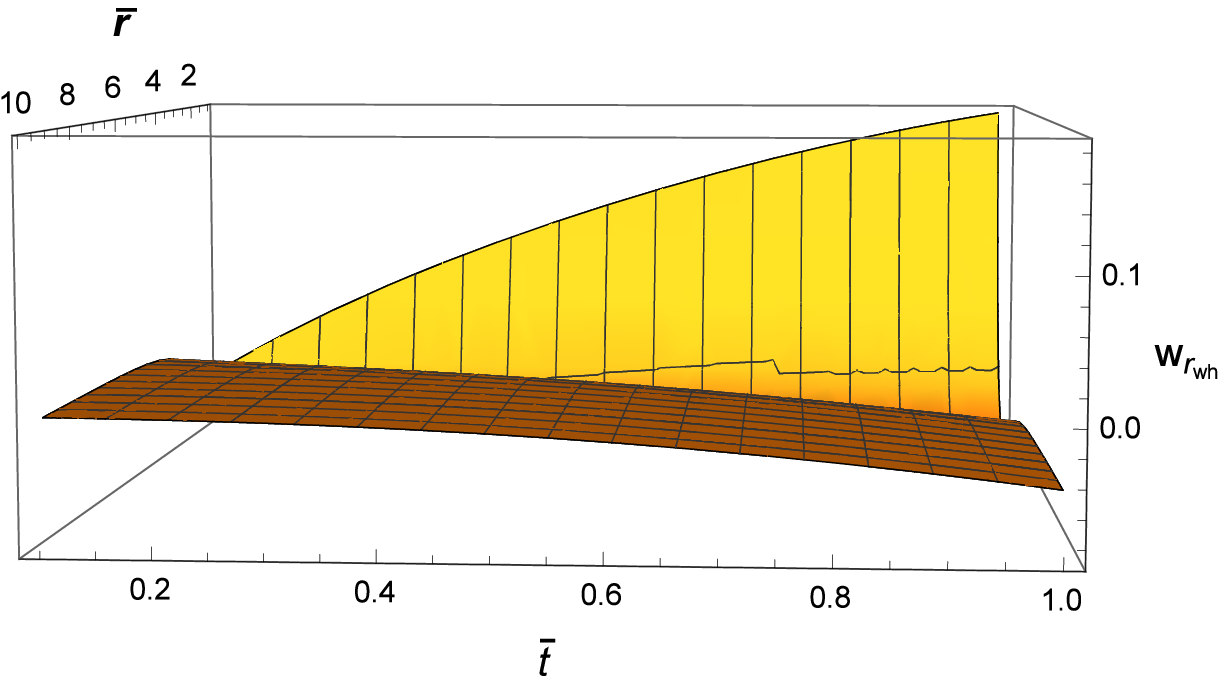}\label{Fig13}}
    \hfill
    \subfloat[Exoticity]{\includegraphics[width=0.45\textwidth]{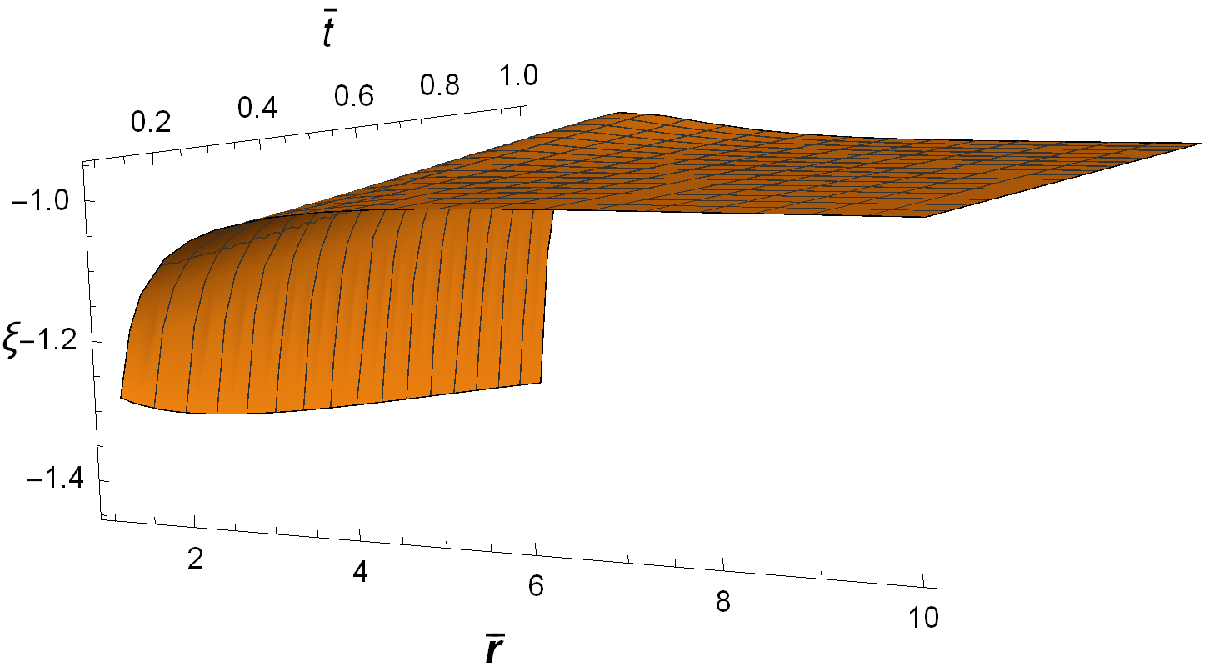}\label{Fig13b}}
    \caption{\scriptsize{Figures show the wormhole state parameter $w_{r_{wh}}=p_{r_{wh}}/\rho_{r_{wh}}$ and the
        exoticity parameter $\xi=(p-\rho)/|\rho|$ for matter dominated era as a function of the
        dimensionless time $\bar{t}=t/t_{0}$ and the dimensionless radius $\bar{r}=r/r_{0}$ . T
        he parameters used were $m=2$, $n=2/3$, $\Lambda=0.001$, $\varphi_{0}=\zeta=1$  
        and  $t/t_{0}=0.1$ as the origin with a critical density $\rho_{c}=3/(\kappa t_{0}^2)$, the
        shape function $b(r)=r_{0}^{m+1}/r^{m}$, the redshift function
        given by $\Phi(t,r)=\varphi_{0}e^{(-r/r_{0}+1)^n}\phi(t)$ (with $\phi(t)$ being displayed in Fig. (\ref{Fig9}) and the function $f(R)$ given by (\ref{fmatter}))}}
\end{figure}

While considering the matter dominated era, characterized with $w=0$ and $a(t)=a_{0}(t/t_{0})^{n}$, we again analyse the case of small dynamical wormhole, $r_{c}=10 \times r_{0}$,
with shape and red-shift functions given by $b(r)=r_{0}^{m+1}/r^{m}$ and $h(r)= \varphi_{0} e^{(-r/r_{0}+1)^{n}}$ respectively. We also set $\zeta=1$. We are interested
in the evolution of dynamic wormhole solutions from the beginning of matter domination era, $t/t_{0}=10^{-6}$, until now, $t/t_{0}=1$. Using the same approach as
above to solve the non-diagonal equation for $\phi(t)$ (see Fig.~\ref{Fig9}), we compute the components of
energy momentum tensor for matter supporting the considered cosmological wormhole. We depict the time evolution and space dependence of energy density, as well as the sum of energy density and both radial and lateral pressures in Fig~\ref{Fig10}, Fig.~\ref{Fig11} and Fig.~\ref{Fig12} respectively. As similar to the solutions in the radiation domination phase,
there is no WEC violation present for the supporting matter during the matter dominated era. In Fig.~\ref{Fig13b} is sketched the exoticity parameter for the matter era. We can notice a similar behaviour as the previous section since $\xi$ is always negative. Thus, the matter remains non-exotic and hence the wormhole is stable.

\begin{figure}[H]
\centering  \includegraphics[width=0.8\textwidth]{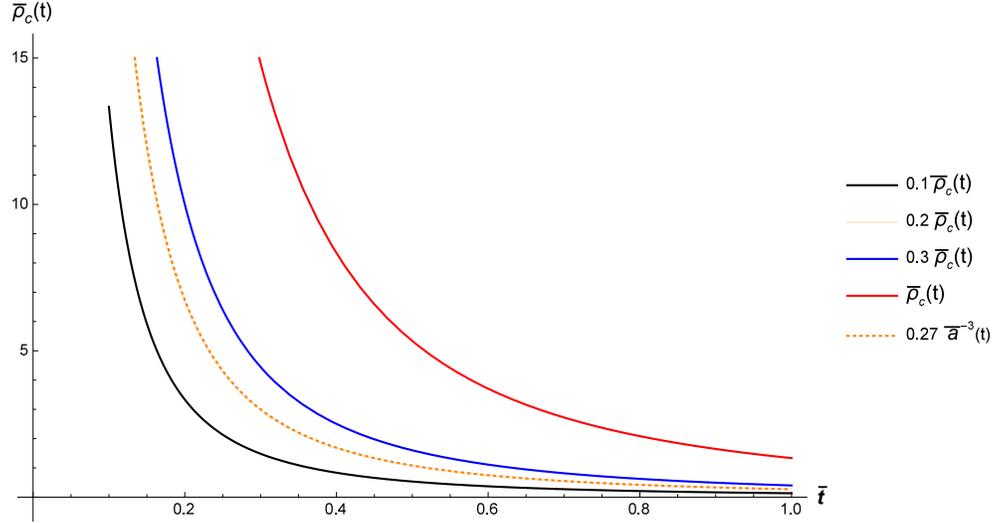}
    \caption{\scriptsize{Figure shows a family of curves with the same evolution as the cosmological dimensionless density  $\bar{\rho}_{c}(t)$ evaluated at $r_{c}=10$ versus dimensionless
        time $\bar{t}=t/t_{0}$. The curve $0.27\bar{a}(t)^{-3}=0.27\bar{t}^{-2}$  is overlaped with the curve $0.2\,\bar{\rho}_{c}(t)$,
        so that  $C=0.2$ is a suitable value.
    The parameters used were $m=2$, $n=2/3$, $\Lambda=0.001$, $\varphi_{0}=\zeta=1$,
        and  $t/t_{0}=0.1$ as the origin with a critical density $\rho_{c}=3/(\kappa t_{0}^2)$, a shape function $b(r)=r_{0}^{m+1}/r^{m}$, a redshift function given by $\Phi(t,r)=\varphi_{0}e^{(-r/r_{0}+1)^n}\phi(t)$ (with $\phi(t)$ being displayed in Fig. (\ref{Fig9}) and the function $f(R)$ given by (\ref{fmatter})).}}\label{Fig16}
\end{figure}

\begin{figure}[H]
    \centering  \includegraphics[width=0.5\textwidth]{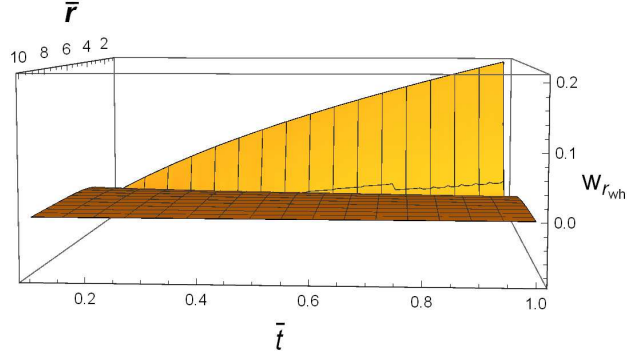}
    \caption{\scriptsize{Wormhole state parameter $w_{r_{wh}}=p_{r_{wh}}/\rho_{r_{wh}}$ for matter dominated era with $\Lambda=0$ as a function of the
        dimensionless time $\bar{t}=t/t_{0}$ and the dimensionless radius $\bar{r}=r/r_{0}$ . T
        he parameters used were $m=2$, $n=2/3$, $\varphi_{0}=\zeta=1$  
        and  $t/t_{0}=0.1$ as the origin with a critical density $\rho_{c}=3/(\kappa t_{0}^2)$, the
        shape function $b(r)=r_{0}^{m+1}/r^{m}$, the redshift function
        given by $\Phi(t,r)=\varphi_{0}e^{(-r/r_{0}+1)^n}\phi(t)$ (with $\phi(t)$ being displayed in Fig. (\ref{Fig9}) and the function $f(R)$ given by (\ref{fmatter}))}}\label{Fig14}
\end{figure}

From the plot for time evolution of radial EOS parameter, $ w_{r_{wh}}=p_{r_{wh}}/\rho_{wh}$, shown in Fig.~\ref{Fig13}, one can see that it at first approaches $w_{r_{wh}}=0$, but then goes to the negative values. This comes from the fact that constant term
in the $f(R)$ expansion, which effectively describes the cosmological constant, starts to dominate over the matter when approaching $t_{0}$. This is, of course, completely consistent with our current understanding
of the cosmological evolution. We demonstrate this with showing $ w_{r_{wh}}$ for the case when $\Lambda=0$ (see Fig.~\ref{Fig14}), where it is clear that the wormhole EOS goes to the
cosmological one when there is only matter present. Result depicted in Fig.~\ref{Fig15} furthermore shows that the difference between radial and lateral pressure
goes to zero, and therefore wormhole supporting matter approaches the ideal cosmological fluid as the radial coordinate
goes to $r_{c}$, so that it can be smoothly matched with the FLRW universe for $r>r_{c}$. Finally, in the same manner as for the radiation domination era, we demonstrate that the boundary condition $\rho_{wh}(r_{c},t)=\rho_{c}(t)$ can be exactly
numerically satisfied in the Fig.~\ref{Fig16}.


\subsection{$f(R)=R$ limit}
\begin{figure}[H]
    \centering
    \subfloat[WEC-1]{\includegraphics[width=0.5\textwidth]{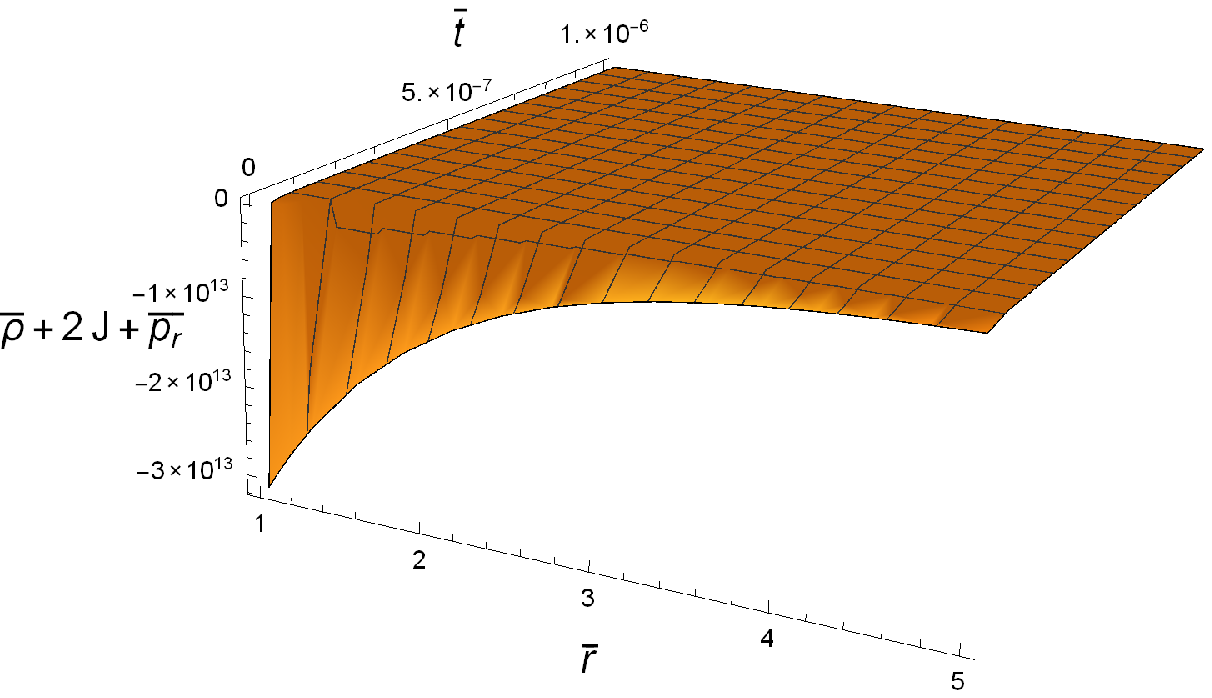}\label{Fig2GR}}
    \hfill
    \subfloat[WEC-2]{\includegraphics[width=0.5\textwidth]{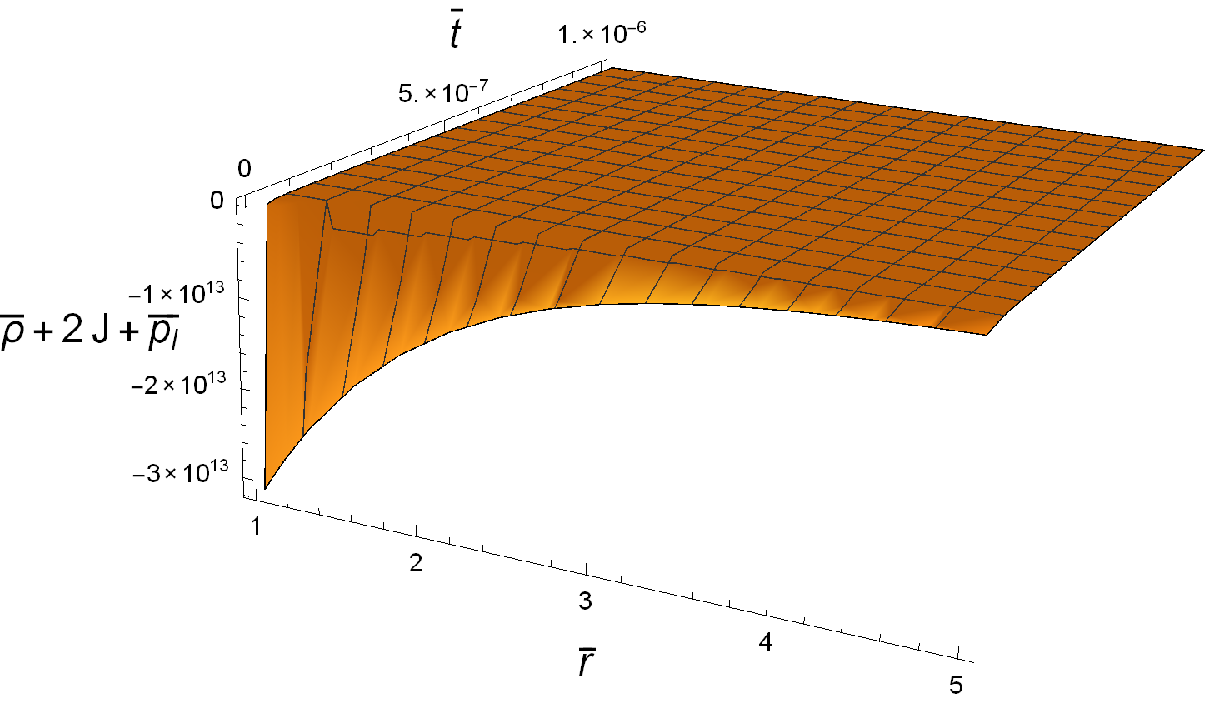}\label{Fig3GR}}
    \hfill
    \caption{\small{Figures show WEC-1 ($\bar{p}_{r}+\bar{\rho}+2J$) and WEC-2 ($\bar{p}_{l}+\bar{\rho}+2J$) for radiation era as a function of the dimensionless time $\bar{t}=t/t_{0}$ and the dimensionless radius $\bar{r}=r/r_{0}$  respectively. The parameters used were $m=2$, $n=1/2$, 
         and $t/t_{0}=10^{-9}$ as the origin with a critical density $\rho_{c}=3/(\kappa t_{0}^2)$, a shape function $b(r)=r_{0}^{m+1}/r^{m}$, a redshift function given by $\Phi(r,t)=a(t)^{-1}\varphi_{0}e^{(-r/r_{0}+1)^n}$  and the function $f(R)=R$.}}
\end{figure}

\begin{figure}[H]
    \centering
    \subfloat[WEC-1]{\includegraphics[width=0.5\textwidth]{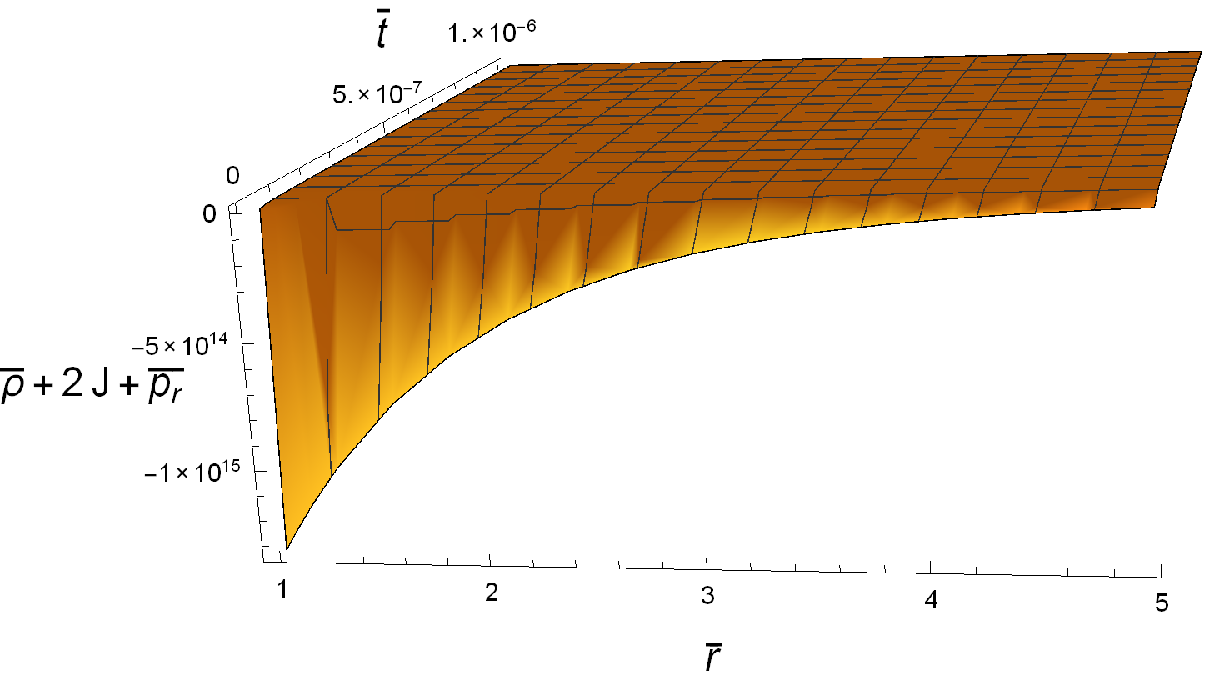}\label{Fig6GR}}
    \hfill
    \subfloat[WEC-2]{\includegraphics[width=0.5\textwidth]{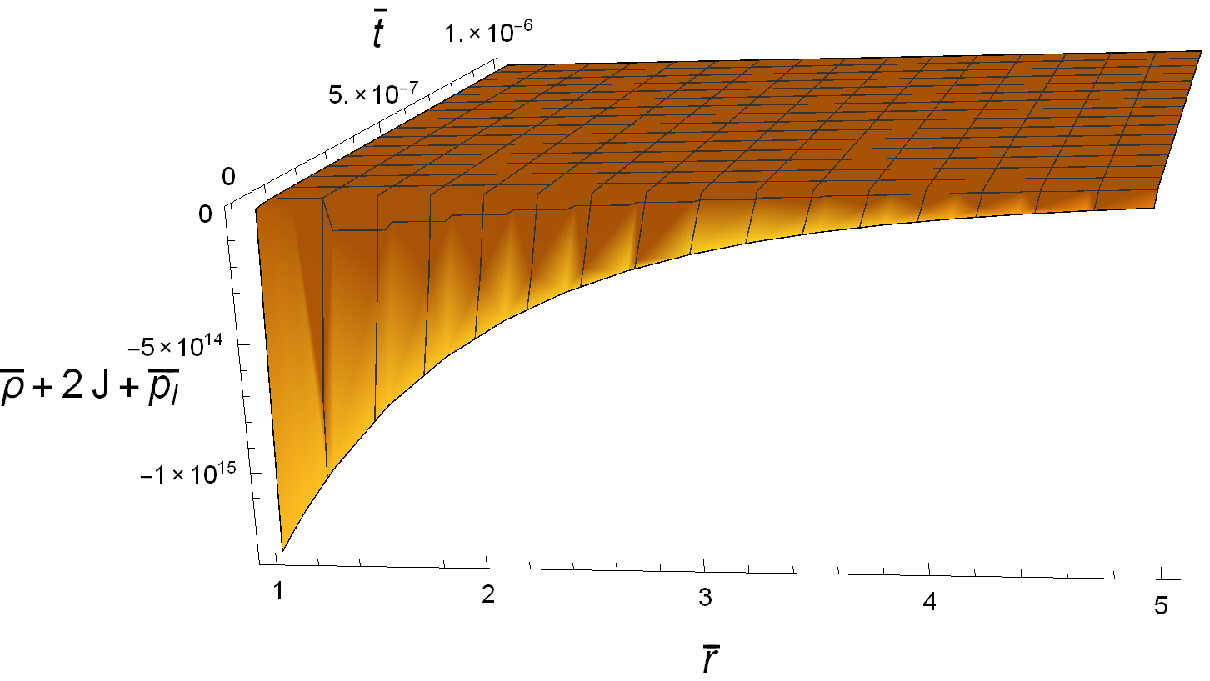}\label{Fig7GR}}
    \hfill
    \caption{\small{Figures show WEC-1 ($\bar{p}_{r}+\bar{\rho}+2J$) and WEC-2  ($\bar{p}_{l}+\bar{\rho}+2J$) for matter era as a function of the dimensionless time $\bar{t}=t/t_{0}$ and dimensionless radius $\bar{r}=r/r_{0}$ respectively. In this case, we have used the parameters $m=2$, $n=2/3$
        and  $t/t_{0}=0.1$ as the origin with a critical density $\rho_{c}=3/(\kappa t_{0}^2)$, a shape function $b(r)=r_{0}^{m+1}/r^{m}$, a redshift function given by $\Phi(t,r)=a^{-1}(t)\varphi_{0}e^{(-r/r_{0}+1)^n}$ and the function $f(R)=R$} }
\end{figure}

A case of special interest in the theoretical framework we use is given by the limit of Einstein's GR, $f(R)=R$. As was shown in \cite{Hochberg}, in general WEC is necessary violated near the throat in time-dependent wormholes.
This is also confirmed in the cosmological solutions we obtain. Taking the special case of standard GR, wormhole solutions cannot in general be supported by considered anisotropic ideal fluid
represented by only diagonal components of energy-momentum tensor. In order to solve the off-diagonal field equation (\ref{fieldeq1d})
in the case of general dynamic wormhole, one needs to introduce additional off-diagonal component of anisotropic fluid, $T^{t}_{r}=J(r,t)$, which
corresponds to the energy flux of the fluid. Off-diagonal equation now reads:
\begin{equation}
2\big(\frac{\dot{a}}{a}\big)\Phi'=J(r,t)\, .
\end{equation}
This means that the WEC, $T^{\mu \nu}k_{\mu} k_{\nu} \geq 0$, will now also include this energy flux:
\begin{equation}
\rho_{wh}(r,t) \geq 0\,,
\end{equation}
\begin{equation}
\rho_{wh}(r,t) + p_{r_{wh}}(r,t) + 2J(r,t) \geq 0\,,
\end{equation}
\begin{equation}
\rho_{wh}(r,t) + p_{l_{wh}}(r,t) + 2J(r,t) \geq 0 \,.
\end{equation}
We again take $h(r)= \varphi_{0} e^{(-r/r_{0}+1)^{n}}$, but now we also have the freedom to prescribe $\phi(t)$.
Since $a(t)$ defines the natural time scale of the system under consideration it seems natural to take
$\phi(t)=a_{0}/a(t)$.  We show solutions of the field equations for
$f(R)=R$ case, for radiation dominated universe in Figs.~\ref{Fig2GR}$-$\ref{Fig3GR}, and for the matter dominated universe in Figs.~\ref{Fig6GR}$-$\ref{Fig7GR}.
One can see that in the case of standard General relativity our solutions
lead to the WEC violation. Therefore, an important difference between
Einstein's GR and cosmologically viable reconstructed $f(R)$ theory lies in the fact that only later admits wormhole
solution supported by regular matter during the whole period of matter and radiation domination.

\section{Discussion}

Discussion on the possible existence of wormholes as astrophysical objects naturally leads to the question of their properties
and evolution during the expansion of the universe. We speculate that cosmological wormholes could be created in the conditions
characterizing the early universe plasma, and then subsequently evolve during the radiation and matter dominance epoch. In
order to describe wormholes in a cosmological context it is necessary to match evolving wormhole geometry with the FLRW space-time.
In the present work this was achieved by using the approximation of a small wormhole - where geometry is determined by requirement
that for some radial distance away from the throat of the wormhole, $r_{c}$, shape and red-shift functions become so small that they can be taken to
vanish, therefore enabling wormhole matching to the FLRW geometry. When approaching $r_{c}$, anisotropic fluid supporting the wormhole needs to go to the ideal isotropic fluid of the universe, with equation of state parameter going to the one characterizing radiation and matter
dominance epoch respectively.
\\
\\
Our analysis was done in the framework of $f(R)$ modified gravity, which is of special interest for several reasons: it represents a simple and natural mathematical generalization of standard GR, it does not introduce any new physical assumptions or entities, it is
capable of describing observed accelerated expansion of the universe, and avoiding singularities which appear in the standard GR.
Interested in viable cosmological solutions we choose the form of $(R)$ functions given by the reconstruction procedure,
which leads to the known evolution of cosmological scale factor. We considered the wormhole solutions that were described by the same scale factor as the universe -
i.e. that had its dynamics determined by the expansion of the universe. Choosing simple functions for shape parameter and space-dependent
part of red-shift function we constructed examples of cosmological wormhole solutions in radiation and matter dominance epoch.
It was demonstrated that there is no WEC violation for material supporting these wormholes,
while in the $f(R)=R$ limit WEC will always be violated
in both epochs.
Such cosmological wormholes could then be created as a microscopic objects in the early universe,
and then increase their size during the evolution of the universe, until they reach the size of
average astrophysical objects today.
\\
\\
In this paper, we have focused on the evolution of wormhole in the radiation and dominated phases of the universe.  Working withing the $f(R)$ framework
 we assume that dark energy is effectively described by modification of the action for the GR.
 If the universe is dominated by some form of exotic matter such as phantom energy, than phantom wormholes within $f(R)$ theory would be of theoretical interest as well.
\\
\\
Presented investigation of cosmological wormholes opens many potential questions for further work.
It would be important to extend the analysis to the era of inflation and late time expansion,
described by the scale factor exponentially dependent on time. Another interesting direction would be to look for
evolving wormhole construction within the curvature-matter coupling models with actions such as $f(R)\mathcal{L}_m$ or $f(R,T)$, where $T$ is
the trace of the energy-momentum tensor. Moreover thermodynamic aspects of such wormholes at their apparent or event horizons along with further
extensions to higher dimensions would be of some interest too. Feature of special physical interest would be to discuss astrophysical properties and possibility
of detection for discussed solutions, as well as their interaction with surrounding matter - for instance via
accretion of different material on the cosmological wormholes.

\section{Acknowledgment}
S.B. is supported by the Comisi\'on Nacional de Investigaci\'on
Cient\'ifica y Tecnol\'ogica (Becas Chile Grant No. 72150066).


\begin{thebibliography}{99}

    \bibitem{moris} M.S. Morris, K.S. Thorne, Am. J. Phys. 56, 395 (1988).

        \bibitem{Flamm} L. Flamm ``Beitraege zur Einsteinischen Gritationstheorie". Physikalische Zeitscrift XVII: 448 (1916)

        \bibitem{Einstein} A. Einstein, N. Rosen, Phys. Rev. 48, 73 (1935)

    \bibitem{phantom} M. Jamil, M.U. Farooq, M.A. Rashid, Eur. Phys. J. C 59, 907 (2009); M. Jamil, P.K.F. Kuhfittig, F. Rahaman, Sk. A Rakib, Eur. Phys. J. C 67, 513 (2010).

    \bibitem{reviews}  A. De Felice, S. Tsujikawa, Living Rev. Rel. 13: 3, (2010);
     S. Nojiri, S. Odintsov, Int. J. Geom. Meth. Mod. Phys. 4, 115 (2007); ibid, Phys. Rept. 505, 59 (2011);
I.~de Martino, M.~De Laurentis and S.~Capozziello,
Universe 1, 123 (2015); K. Bamba, S. Capozziello, S. Nojiri, S.
Odintsov, Astrophys. Space Sci. 342, 155 (2012); S. Capozziello, M.
De Laurentis, arXiv:1108.6266

    \bibitem{harko}  T. Harko, F. S. N. Lobo, M. K. Mak, S. V. Sushkov, Phys. Rev. D 87, 067504 (2013); F.S.N. Lobo, AIP Conf. Proc. 1458, 447 (2011); F.S.N. Lobo, M. A. Oliveira, Phys. Rev. D 80, 104012 (2009)

    \bibitem{fur} N. Furey, A. De Benedictis, Class. Quant. Grav. 22, 313 (2005); A. De Benedictis, D. Horvat, Gen. Rel. Grav. 44, 2711 (2012).

    \bibitem{r} F. Rahaman, A. Banerjee, M. Jamil, A. K. Yadav, H. Idris,  Int. J. Theor. Phys. 53, 1910 (2014);
     M. Jamil, F. Rahaman, R. Myrzakulov, P.K.F. Kuhfittig, N. Ahmed, U.F. Mondal, J. Korean Phys. Soc. 65, 917 (2014).

        \bibitem{viable} P. Pavlovic, M. Sossich, Eur.Phys.J. C75 117 (2015).

\bibitem{Myrzakulov:2015kda}
R.~Myrzakulov, L.~Sebastiani, S.~Vagnozzi and S.~Zerbini,
Class.\ Quant.\ Grav.\  {\bf 33} no.12,  125005 (2016).

  \bibitem{azizi} T. Azizi, Int. J. Theo. Phys. 52, 3486 (2013); N.M. Garcia, F. S. N. Lobo, Phys. Rev. D 82, 104018 (2010).

    \bibitem{tg}  C.G. Boehmer, T. Harko, F.S.N. Lobo, Phys. Rev. D 85, 044033 (2012);  M. Jamil, D. Momeni,
     R. Myrzakulov, Eur. Phys. J. C 73, 2267 (2013).

    \bibitem{gb} Z. Amirabi, M. Halilsoy, S.H. Mazharimousavi, Phys. Rev. D 88, 124023 (2013).

         \bibitem{lbd} D. W. Tian arXiv:1508.02291 [gr-qc]
         
         \bibitem{Ayon-Beato:2015eca}
         E.~Ayon-Beato, F.~Canfora and J.~Zanelli,
         Phys.\ Lett.\ B 752, 201 (2016).
         
    \bibitem{trobo} L. A. Anchordoqui, D. F. Torres, M. L. Trobo, S. E. P. Bergliaffa, Phys. Rev. D 57, 829 (1998);
    S.~W.~Kim,
    Phys.\ Rev.\ D 53, 6889 (1996);
    M.~Cataldo, S.~del Campo, P.~Minning and P.~Salgado,
    Phys.\ Rev.\ D  79, 024005 (2009);
    M.~Cataldo and S.~del Campo,
    Phys.\ Rev.\ D 85, 104010 (2012);
M.~Cataldo, P.~Meza and P.~Minning,
Phys.\ Rev.\ D 83, 044050 (2011);
    M.~Cataldo, F.~Ar\'ostica and S.~Bahamonde,
    Eur.\ Phys.\ J.\ C 73, 2517 (2013);
    M.~Cataldo, F.~Arostica and S.~Bahamonde,
    Phys.\ Rev.\ D 88, 047502 (2013);
    U.~Debnath, M.~Jamil, R.~Myrzakulov and M.~Akbar,
    Int.\ J.\ Theor.\ Phys.\  53, 4083 (2014).
    M.~U.~Farooq, M.~Akbar and M.~Jamil,
    AIP Conf.\ Proc.\  1295, 176 (2010).

    \bibitem{teo} E. Teo, Phys. Rev. D 58, 024014 (1998); P. K. F. Kuhfittig, Phys. Rev. D 67, 064015
    (2003).

        \bibitem{Hochberg} D. Hochberg, Matt Visser, Phys. Rev. Lett. 81, 746
        (1998).

        \bibitem{saeidi} H. Saeidi and B. N. Esfahani, Mod. Phys. Lett. A 26, 1211
        (2011).

        \bibitem{sharif} M. Sharif, S. Rani, Gen. Rel. Grav. 45, 2389
        (2013).

        \bibitem{Rahaman} F. Rahaman, P.K.F. Kuhfittig, S. Ray, N. Islam, Eur. Phys. J. C 74, 2750
        (2014).

        \bibitem{kuh} P. K. Kuhfittig  Eur. Phys. J. C 74, 2818 (2014); Z. L., C. Bambi, Phys. Rev. D 90, 024071 (2014)
\bibitem{haw}
S.~W.~Hawking and G.~F.~R.~Ellis,
``The Large Scale Structure of Space-Time,'' (Vol. 1) Cambridge University Press 1973
doi:10.1017/CBO9780511524646


        \bibitem{hideki} H. Maeda, T. Harada, B.J. Carr, Phys. Rev. D
        79, 044034 (2009).

        \bibitem{reconstruction} J. He, B. Wang, Phys. Rev. D 87,  023508
        (2013).

\bibitem{Nojiri:2006gh}
S.~Nojiri and S.~D.~Odintsov,
Phys.\ Rev.\ D 74, 086005 (2006).

\bibitem{Nojiri:2009kx}
S.~Nojiri, S.~D.~Odintsov and D.~Saez-Gomez,
Phys.\ Lett.\ B 681, 74 (2009).


\bibitem{Nojiri:2006be}
S.~Nojiri and S.~D.~Odintsov,
J.\ Phys.\ Conf.\ Ser.\   66, 012005 (2007).

\bibitem{Dunsby:2010wg}
P.~K.~S.~Dunsby, E.~Elizalde, R.~Goswami, S.~Odintsov and D.~S.~Gomez,
Phys.\ Rev.\ D  82, 023519 (2010).


        \bibitem{sawicki} I. Sawicki, W. Hu, Phys. Rev. D 75, 127502
        (2007).

        \bibitem{planck} Planck Collaboration, arXiv:1502.01589 [astro-ph.CO]
        
        \bibitem{Ayon-Beato:2015eca}
        E.~Ayon-Beato, F.~Canfora and J.~Zanelli,
        Phys.\ Lett.\ B 752, 201 (2016).
        
\end{thebibliography}
\end{document}